\documentclass[showpacs,preprint,aps]{revtex4}
\usepackage{graphicx}
\usepackage{dcolumn}
\usepackage{bm}

\begin{document}
\title{Structure of $^{8}$B and astrophysical $S_{17}$ factor
in Skyrme Hartree-Fock theory}

\author{S.S. Chandel and S.K. Dhiman}
\affiliation{Department of Physics, H.P. University, Shimla-171005, India}
\author{R. Shyam}
\affiliation{Saha Institute of Nuclear Physics, AF/1, Bidhan Nagar,
Kolkata -700 064, India.}

\date{\today}

\begin{abstract}
We investigate the ground state structure of $^8$B within the Skyrme
Hartree-Fock framework where spin-orbit part of the effective 
interaction is adjusted to reproduce the one-proton separation energy
of this nucleus. Using same set of force parameters, binding energies
and root mean square radii of other light p-shell unstable nuclei,
$^8$Li, $^7$B, $^7$Be, and $^9$C, have been calculated where a good
agreement with corresponding experimental data is obtained.
The overlap integral of $^8$B
and $^7$Be wave functions has been used to determine the root mean square
radius of the single proton in a particular orbit and also the 
astrophysical S factor ($S_{17}$) for the $^{7}$Be($p, \gamma)^{8}$B
radiative capture reaction. It is found that the asymptotic region
(distances beyond 4 fm) of the p-shell single proton wave function
contributes more than half to the calculated value (4.93 fm) of the
corresponding single particle root mean square radius. The value of   
$S_{17}$ is determined to be  21.1 eV b which is in good agreement
with the recommended value for the near zero energy $S_{17}$ of
$19.0^{+4.0}_{-1.0}$ eV b. 
\end{abstract}
\pacs{ $21.60.J_{z}$, $21.10$, $21.65.+f$, $26.65.+t.$ }
\maketitle

\section{Introduction}
Studies involving rare neutron rich and proton rich isotopes
are currently at the forefront of the nuclear physics research. 
Experiments \cite{tani85,han95} performed with beams of nuclei far from
the beta stability (NFABS) line have revealed many features of these
systems which are not present in their stable counterparts. For example,
some NFABS having very small particle separation energies are
very large in their radial dimensions (they are also called as halo nuclei);
their radii are not governed by the usual $r_0A^{1/3}$ (with r$_0$ = 1.20
and A the mass of the nucleus) rule. The discovery of these nuclei underlines 
the necessity of revising the traditional picture of nuclear
structure in important ways since away from the $\beta$-stability
nuclear dynamics are characterized by a variety of new features not
present in stable nuclei. In the halo region the quantal dynamical effects
play a crucial role in distribution of the nuclear density
in the zone of very weak binding.  

Proper knowledge of the structure of $^{8}$B nucleus is   
important for several reasons. This nucleus perhaps is 
the most likely candidate for having a proton halo structure,
as its last proton has a binding energy of only 137 keV. 
$^8$B produced via $^7$Be(p,$\gamma)^8$B reaction in the
Sun is the source of high energy neutrinos which are detected
in SNO, Kamiokande and Homestake experiments \cite{bac89,dav94,fuk01}. 
Therefore, accurate determination of the cross
section of this reaction at relative energies
corresponding to solar temperatures (about 20 keV) is very important 
to the solar neutrino issue. In this energy region, the cross-section
$\sigma_{p \gamma}(E_{cm})$ [which is usually expressed in terms of
the astrophysical $S_{17}(E_{cm})$ factor] of the $^7$Be(p,$\gamma)^8$B
capture reaction is directly proportional to the high energy solar
neutrino flux.  A better knowledge of $S_{17}$ is, therefore,
important to improve the precision of the theoretical prediction of
$^8$B neutrino flux from present and future solar neutrino experiments.

$^{7}$Be (p,$\gamma)^{8}$B reaction has been studied extensively both
theoretically as well as experimentally
\cite{tani88,neg96,bla97,esb00,bar97,shy98,tim98,car01,sim99,gri98,obu96}.
$S_{17}$ is determined
either by direct measurements \cite{fil83} or by indirect methods
such as Coulomb dissociation \cite{mot94} of $^8$B on heavy targets
and transfer reactions 
in which $^8$B is either the residual nucleus or the projectile nucleus 
\cite{xu94,gag01,azh99}. Efforts have also been made to calculate the
cross section of this reaction within the framework of the shell model
and the cluster model \cite{tim97,cso00,bro96}.
The key point of these calculations is the determination of the 
wave functions of $^8$B states within the given structure theory.
 
Aim of the present study is to investigate the structure of $^8$B
in the framework of the Skyrme Hartree-Fock (SkHF) model which has
been used successfully to describe the ground-state properties of
both stable \cite{neg72,fri86,dob94} as well as exotic nuclei 
\cite{yao96,li96,miz00,rei99}.
The phenomena of nuclear skin and halo
have been studied in medium mass and heavy nuclei \cite{miz00,rei99} within
the Skyrme-Hartree-Fock-Bogoliubov method using the SLy4 Skyrme force.
The SkHF method with density-dependent pairing correlation and SkI4
interaction parameters has been successful in reproducing the binding
energies and rms radii \cite{miz00} in the light neutron halo nuclei
$^6$He, $^8$He, $^{11}$Li and $^{14}$Be.

We solve axially symmetric Hartree-Fock (HF) equations with SLy4
\cite{cha97} Skyrme interaction which has been
constructed by fitting to the experimental data on
radii and binding energies of symmetric and neutron-rich nuclei.
This has also been used in Ref. \cite{rei99} to study the phenomenon of
shape coexistence in semi-magic isotopes of Mg, S and Zr nuclei.
In our calculations pairing correlations among nucleons have been
treated within the BCS pairing method. We have, however,
renormalized the parameter of the spin-orbit term of the SLy4
interaction so as to reproduce the experimental binding energy of the last
proton in $^{8}$B nucleus. A check on our interaction parameters was
made  by calculating binding energies and rms radii of $^7$Be, $^7$B,
$^8$Li and $^9$C nuclei with the same set where a 
good agreement is obtained with corresponding experimental data.
We calculate the root mean square radii for matter, neutron and 
proton distributions for $^8$B. Using the overlap integral of  
HF wave functions for $^7$Be and $^8$B ground states, 
the root mean square radius of the valence p-shell proton in this 
nucleus has been determined which is expected to provide information
about the proton halo structure in $^8$B. The overlap integral has also
been  used to calculate the astrophysical $S_{17}$ factor.
 
The paper is organized in the following way. In section II we present
the short description of the method of our calculations.  Section III
contains our results and their discussions.  Summary and  
conclusions of  our study are given in section IV.

\section{METHOD OF CALCULATIONS}

In the Skyrme Hartree-Fock formalism, binding energies, densities and
single particle wave functions are obtained from a local energy functional.
Skyrme force parameters are determined empirically
by fitting nuclear matter properties of stable nuclei and neutron star 
densities \cite{cha97}. Microscopically, the
Skyrme functional corresponds to an 
expansion of the nuclear interaction up to the first order in momentum 
transfer \cite{neg72}. The results for ground state properties are derived 
self-consistently from the total energy functional of the nucleus
which is given by
\begin{eqnarray}
E = E_{KE}+E_{sk}(\rho,\tau)+E_{sk,ls}(\rho,J)+E_{Coul}(\rho_p)
{} \nonumber\\
+E_{Coul,exch}(\rho_{p})+E_{pair}-E_{CM}
\end{eqnarray}
where $\rho_p$ is the proton density and
\begin{equation}
E_{KE} = 4\pi \sum_{q\in p,n}\int_{0}^{\infty}dr
 r^{2}\frac{\hbar}{2m_{q}}\tau_{q},
\end{equation}
\begin{eqnarray}
E_{Sk}(\rho,\tau) = 4\pi \int_{0}^{\infty} drr^2 \left( \frac{b_0}{2.}\rho^2
+ b_1\rho\tau - \frac{b_2}{2.}\rho\nabla^2\rho + \frac{b_3}{3.}
\rho^{\beta+2.}\right) {}  \nonumber \\
- 4\pi\sum_{q\in p,n } \int_{0}^{\infty}drr^2 \left( \frac{b^{^{\prime}}_0}{2.}
\rho^2_q + b^{^{\prime}}_1\rho_q\tau_q - \frac{b^{^{\prime}}_2}{2.} \rho_{q}
\nabla^{2}\rho_{q} -\frac{b^{^{\prime}}_3}{3.} \rho^{\beta}
\rho^{2}_{q}\right),
\end{eqnarray}
\begin{equation}
E_{Sk,ls}(\rho,J) = -4\pi\int_{0}^{\infty}dr r^2\left( b_4\rho \nabla J +
b^{^{\prime}}_4 \rho_{q} \nabla J_q \right).
\end{equation}
In above equations $q\in\left\{ p,n \right\}$ is the isospin label
for proton and neutron.
In Eq. (4) an additional coefficient $b^{^{\prime}}_{4}$ has been introduced in
the spin-orbit term of the energy functional. The Coulomb energy term is given by 
\begin{equation}
E_{Coul}(\rho_p) = \frac{1}{2} e^2\int d^3r d^3r^{^{\prime}} \rho_C(r)
\frac{1.} {\mid r- r^{^{\prime}}\mid} \rho_C(r^{^{\prime}}).
\end{equation}
$E_{Coul, exch}$ is Coulomb exchange energy functional and it is
usually treated in the Slater approximation,
\begin{equation}
E_{Coul,exch}(\rho_p) = -\frac{3}{4} (\frac{3}{\pi})^{1/3} 4\pi
\int_0^{\infty} dr r^2 \rho^{\frac{4}{3}}_p(r).
\end{equation}
The center of mass energy functional $E_{CM}$ is written as,
\begin{equation}
E_{CM} =\frac{\langle P^2_{CM}\rangle}{2Am},
\end{equation}
where $P_{CM} = \sum_i \hat{p}$ is the total angular momentum operator
in the center of mass frame, which project out a state with good total
angular momentum in the given mean-field state.
We account for the pairing interaction among nucleons by solving BCS 
equations. The schematic pairing energy functional is given as
\begin{equation}
E_{pair} = -\sum_{q\in p,n} G_q \left[ \sum_{\alpha \in q} \sqrt{%
v_{\alpha}(1 - v_{\alpha})} \right]^2.
\end{equation}
The pairing strength parameters are $G_p = 29/$A MeV for proton and
$G_n = 29/$A MeV for neutron where A is number of nucleons in the nucleus.
The occupation probabilities are given by 
\begin{equation}
v_{\alpha}^2 = \frac{1}{2} \left[ 1-\frac{(\epsilon_{\alpha} -
\epsilon_{F,\alpha})^2}{(\epsilon_{\alpha} - \epsilon_{F,\alpha})^2 +
\Delta^2_q} \right],
\end{equation}
where $\epsilon_\alpha$ is the single particle energy of the given orbit.
The pairing gap $\Delta_q $ and the Fermi energy $\epsilon_{F,q}$ are
obtained by the iterative solution of the gap equation and the particle number
condition. Axially symmetric single particle wave functions for a nucleus
are given in expansion basis by
\begin{equation} 
\psi_{\alpha}(r) = \phi_{\alpha}(r)\times
Y_{l_{\alpha}j_{\alpha}m_{\alpha}}(\theta,\phi),
\end{equation} 
where $\alpha$ represents the quantum numbers of the state $\mid nljm>$
and spinor-spherical harmonics are given by
$Y_{l_{\alpha}j_{\alpha}m_{\alpha}}(\theta, \phi)$. 
$\phi_{\alpha}(r)$ represents the radial wave function. 
Various phenomenological densities are used to construct the energy
functional, particle densities $\rho_q(r) = \sum_{\alpha\in
q}N_{\alpha}\mid \psi_{\alpha}(r)\mid^2$, kinetic densities $\tau_q(r) =
\sum_{q\in q}N_{\alpha}\mid \nabla\psi_{\alpha}(r)\mid^2$, and 
spin-orbit densities $\nabla J_q(r) = -i\sum_{\alpha \in q}N_\alpha \nabla
\psi_{\alpha}(r)^{+} \cdot \nabla \times \sigma \psi_{\alpha}(r)$. 
Usual total isoscalar densities without an isospin
label are defined as $\rho = \rho_{\pi} + \rho_{\nu}$,
$\tau = \tau_{\pi} + \tau_{\nu}$ and $J = J_{ \pi} + J_{\nu}$, with
summations being carried out over both proton and neutron
particle numbers. $N_{\alpha}$ represents the desired proton or neutron
number and is equal to $v_{\alpha}^2$ which is the BCS occupation
probabilities of orbitals.

\section{Results and discussions}

\subsection{Binding energies, radii and density distributions}
 
The values of various parameters of the $SLy4$ Skyrme
effective interaction as used in our calculations are: 
$t_0 = -2488.91$, $t_1 = 486.82$, $t_2 = -546.39$, $t_3 = 13777.0$,
$b_4 = 61.5$, $x_0 = 0.83$, $x_1 = -0.34$, $x_2 = -1.00$,
$x_3 = 1.35$, $\beta = \frac{1}{6}.$ The values of $b_{i}$ and
$b_i^{\prime}$ parameters in Eqs. (3) and (4)  have been obtained 
from $t_{i}$ and $x_{i}$ by using the relations given in the Appendix A
of Ref. \cite{rei99}.  We use two-body zero-range spin-orbit interaction 
by taking $b_4^{\prime} = b_4$ combination \cite{fri86}.  
These parameters are the same as those given in Refs. \cite{miz00,rei99}
except for the  spin-orbit term which has been renormalized by taking   
$b_4 = 37.42 $ so as to reproduce the single proton separation energy 
of $^8$B. The adjusted values of the parameter $b_4$ is smaller than
its original value of 61.50. This observation is in line with the
weakening of the spin-orbit interaction in light drip line nuclei noted
in Ref. \cite{lal98}. In the following the force with renormalized 
$b_4$ will be referred to as TH1 and that with the original $b_4$ as 
TH2. 

With the same set of force parameters we have calculated total binding
energies and root mean square (rms) radii of light unstable nuclei,
$^8$Li, $^8$B, $^7$B, $^7$Be, and $^9$C. Results for
binding energies are presented in Table I where the corresponding
experimental values are also shown \cite{fir96}. Binding energies
calculated with TH1 and TH2 forces are shown in third and fourth 
columns of this table, respectively. It is evident that   
with the renormalized $b_4$, our calculations reproduce the 
experimental binding energies of these nuclei to the extent   
of 98$\%$. Furthermore, the single neutron separation energy ($S_n$)
for $^{8}$B as calculated with the same force parameters is  
$13.47$ MeV which is in good agreement with the  corresponding
experimental value $13.02$ MeV.

The rms radii for matter ($r_m$), neutron ($r_n$) and proton ($r_p$)
distributions are presented in Table II for all the five isotopes. Also
shown in the first column of this table are corresponding experimental
values of matter radii ($r_m^{exp}$) \cite{neg96}. $r_m$  
(shown in the second column) has been calculated by
defining the total radius as the average of
proton and neutron radii in every orbit weighted with occupation
probabilities.  $r_p$ and $r_n$ are shown in
third and fourth columns, respectively of this table. We see that
for all the isotopes $r_m$ calculated with the modified $b_4$ 
parameter, is in better agreement with the  
corresponding experimental data as compared to 
that calculated with the original $b_4$. 
 
In Fig. 1, we show distributions of matter, charge, neutron and proton
densities (in the units of fm$^{-3}$) in the coordinate space. 
The nuclear charge density distribution has been obtained by
folding the HF results for proton and neutron densities with
the intrinsic charge density distribution of nucleons
in Fourier-space by transforming the densities to form-factors.
In actual calculations of the nuclear charge density, the 
center of mass correction effects are taken into account by unfolding the 
spurious vibrations of the nuclear center of mass in harmonic 
approximation (as is done in Eq. (7) for the zero-point energy).
We note that nuclear charge and proton densities differ very
slightly from each other. The key point of this figure is that the neutron
and proton densities differ quite a bit from each other for distances
larger than 3 fm, where the proton density develops a long tail. This
is reminiscent of the situation in the neutron halo nuclei where the
neutron density distribution has a long tail \cite{han95}. This observation
supports the existence of a proton halo structure in $^8$B.

\subsection{ Valence proton radius in $^{8}B$ and Astrophysical
$S_{17}$ factor }

To begin our discussions in this section we define the overlap function 
(OF)
\begin{equation}
I_{^{7}Be-p}^{^{8}B}({\bf r})=\left\langle \left[ \psi _{^{7}Be}^{as}\left(
j_{x}\right) \otimes p_{j_{y}}\right] _{JM}\mid \psi _{^{8}B}^{as}\left(
j_{y}\right) \right\rangle _{JM},
\end{equation}
where symbol $\otimes $ denotes vector coupling. In Eq. (11),
J is the total angular momentum which can be obtained by 
projecting the ground state wave function of $^{8}$B  onto 
$^{7}$Be ground state with spin-parity of 3/2$^{-}$.
$\psi_{^{7}Be}$ and $\psi_{^{8}B}$ are SkHF 
wave functions of $^{7}$Be$(p_{3/2})$ and $^{8}$B$\left( j_{y}\right)
(j_{y}=3/2^{-} and 1/2^{-}$). These wave functions have been
obtained with the same force parameters as those described in the
previous section.  

OF for $^{8}B$ nucleus is used in the calculation of the  
asymptotic normalization constant which is related to 
the astrophysical factor $S_{17}$. It is also needed for the calculation
of the valence proton density distribution which could be one more source 
of information about the existent of a proton halo structure in this nucleus.
OF depends upon wave functions of $^{8}$B with given spin-parity
combinations, which are completely antisymmetric in 'n' number
active particles and the core state wave functions with the bound state
having a total angular J and its projection M = 0. Due to the condition
that the norm of OF $(\mid I^{{^8}B}_{{^7}Be-p}\mid)$ is always
$\leq 1$, the overlap function 
is not a wave function of the total Hamiltonian.

The overlap function 
may be expressed in terms of the HF+BCS wave function as
\begin{equation}
I_{^{7}Be-p}^{^{8}B}({\bf r})=\sum_{JM}\langle j_{y}m_{y},j_{x}m_{x}
\mid JM\rangle
Y_{lj_{y},m_{y}}(\theta,\phi)I_{lj_{x},j_{y}}^{^{8}B}(r).
\end{equation}
The radial part of OF is written as,
\begin{equation}
I_{lj_{y}}^{^{8}B}(r)=\sum_{j_{x},j_{y}}v_{j_{x}}v_{j_{y}}
\mid ^{7}Be(j_{x})\otimes p_{j_{y}}\rangle, 
\end{equation}
where $v_{j_x}$ and $v_{j_y}$ are the orbital occupancies of $^7$Be and $^8$B
nuclei, respectively. 
 
The rms radius of the $^{7}$Be + p = $^{8}$B bound state
(or the valence proton) 
can be written in terms of the overlap function as
\begin{equation}
\langle r^{2}\rangle ^{1/2}_{^8B}=\int_0^\infty r^{2}dr
\mid I_{^{7}Be-p}^{^{8}B}(r)\mid.
\end{equation}
The root mean square matter radius can be written in terms  
of $\langle r^2\rangle _{^8B}$ as
\begin{equation}
<r^2_m>_A^{1/2} = \sqrt{\frac{1}{A+1}(A r_c^2 + r^2_p +
 \frac{A}{A+1}\langle r^2\rangle _{^8B})},
\end{equation}
where A is the mass of the core nucleus. 
In Eq. (15), $r_c$ is the rms matter radius of the core (its value is
taken to be $2.33\pm0.02$ fm for $^7Be$), 
$r_p$ is the proton radius (taken to be 0.81 fm).
This definition of
the matter radius assumes that core and the valence particle are 
two distinct clusters and ignores the internal  excitations of the
core nucleus \cite{buc77}. The rms matter radius calculated by
Eq. (15) could be different from that discussed in the previous
section even though HF wave functions have been used in both the
calculations. 
  
In table III, we show results of our calculations for 
$\langle r^{2}\rangle^{1/2}_{^8B}$ and $\langle r^2_m\rangle^{1/2}_A$.
We have also shown in this table corresponding 
results of other authors who have used different models. We note
that our results are in agreement with those of other authors within 
$10-20\% $. Thus, almost all these calculations appear to be in   
agreement with the fact that the valence proton has 
a large spatial extension in $^8$B. One exception, however, is 
Ref. \cite{rii93} where a relatively smaller value of 3.75 fm 
has been reported for the valence proton rms radius in $^8$B. 
These authors have done their calculations by employing
a  two-body model with a Gaussian potential of range 1.90 fm
which may not describe accurately the tail of the p-shell single
proton wave function in $^8$B.

We have estimated the contribution of the asymptotic part of the overlap
wave function to the valence proton rms radius in $^8$B 
in the following way (see, e.g., \cite{car01}). We define the ratio
\begin{equation}
C_\nu(R_N) = \lbrack \int_{R_N}^\infty r^{2\nu}
 dr\mid I_{^7Be-p}^{^8B}(r)\mid^2/\int_0^\infty r^{2\nu} dr
\mid I_{^7Be-p}^{^8B}(r)\mid^2 \rbrack^{1/\nu},
\end{equation}
where $R_N$ is a cut-off radius. $\nu$ = 1 and 2 gives the contribution
of the asymptotic part to the norm of the overlap function and the
rms radius of 
the valence proton, respectively. For $R_N = 2.5$ fm which is the
calculated value of the matter radius in $^7$Be (see table II), we find 
$C_\nu(R_N)$ = 0.6260(0.8579). This indicates that the 
region out side the $^7$Be core contributes up to $85\%$ to the 
valence proton radius in $^8$B and   
that the probability of finding a valence proton outside the $^7$Be core 
nucleus is about $62\%$. For $R_N$ = 4.0 fm (beyond which the nuclear 
interaction becomes negligible) we get $C_{1(2)}(R_N)$ = 0.3248 (0.5650) 
which means that contributions to the valence proton rms radius are 
about $56\%$ from the distances beyond 4 fm. 
These results provide further support to the existence of a 
proton halo structure in $^8$B nucleus.
As remarked earlier, in these calculations we have assumed  
that $^7$Be behaves as an inert core inside the $^8$B nucleus.
Consideration of the excitation of $^7$Be core will not change these
conclusions significantly \cite{car01,mor02}. 

It may be mentioned here that the
experimental value of the quadrupole moment of $^8$B (which is twice
as large as the value predicted by the shell model) can be explained
with a single particle wave function corresponding to a matter radius of 
2.72 fm \cite{neg96,obu96}.
This observation has been thought of as a possible evidence
for a proton halo structure in $^8$B. The matter radius of this nucleus
as calculated by our model is very close to the above value.  
   
Next, we present results for the astrophysical $S_{17}$ factor
calculated within our model.
In the region outside the core, where range of the nuclear interaction
becomes negligible (r $ >$ 4.0 fm), the radial overlap wave function
$I_{lj}^{^{8}B}(r)$ of the bound state can be written as  
\begin{equation}
I_{lj}^{^{8}B}(r)\simeq \bar{c_{lj}}W_{\eta ,l+1/2}(k,r)/r,
\end{equation}
where $W$ is the Whittaker function, $k$ the wave number corresponding
to the single proton separation energy and $\eta $  the Sommerfeld
parameter for the bound state in $^{8}$B. In Eq. (17), 
$\bar{c_{lj}}$ is the asymptotic normalization constant, required
to normalize the Whittaker function to the radial overlap wave function in
the $^{8}$B nucleus in the asymptotic region.
The $S_{17}$ factor is related to the proton capture
cross section as  
\begin{equation}
S_{17}(E)=\sigma (E)E\,e^{(2\pi \eta (E))}.
\end{equation}
It has been shown in ref. \cite{xu94} that at low energies,
the $S_{17}$ factor depends only on $\bar{c_{lj}}$  and one can
write \cite{cso00,bro96} 
\begin{equation}
S_{17}^{A}=C_N\sum_{j_{y}}\bar{c}_{1,j_{y}}^{2},
\end{equation}
The value of $C_N$ depends on the details of the scattering wave
function. In Ref. \cite{cso00}, $C_N$ = 37.8 has been obtained by using 
a microscopic cluster model for the scattering states, while a value of 36.5 
has been reported for this quantity in Ref. \cite{bro96} using a hard-core
scattering state model. 

In our calculations we have used our HF OF directly into
a capture code where the scattering states are described by pure 
Coulomb wave functions between $^7$Be and $p$. Thus, within an 
inert $^7$Be core approximation our results are parameter free.
In Table IV, we show $S_{17}^A$ obtained by using overlap functions
 calculated
with both TH1 and TH2 force parameters. 
We see that $S_{17}^A$ obtained with 
TH1 force is quite close to its 
adopted limits of $19.1^{+4.0}_{-1.0}$ eV b. On the other hand, that
obtained with TH2 is quite large and well beyond the maximum limit
of this value.
 
We also show in this table values of asymptotic normalization 
coefficients and $S_{17}^A$ obtained by using Eq. (19) with $C_N$ =
36.5 and 37.8. It is clear that there is a slight model 
dependence in the $S_{17}$ calculated in this way. These results
however, are in agreement with those obtained by our method within $10\%$.  
 
Alternatively, $S_{17}$ factor can also be written in terms of the
valence proton density around 10 fm ($\rho_{10}$)
and spectroscopic factors of valence proton states as \cite{bro96}:
\begin{equation}
S_{17}^{B}=2.99\times 10^{6}\rho_{10}\times S_{\ell j},
\end{equation}
where $S_{\ell j}=S_{1,3/2}+S_{1,1/2}$ is the total spectroscopic
factor, which may be calculated from the orbital occupancies. 
This relation has been used in Ref. \cite{neg96} to calculate 
$S_{17}$ at the astrophysical energies from the shell model 
spectroscopic factors.
 
In case of $^{8}$B, the last bound proton remains mostly
in $p_{3/2}$ and $p_{1/2}$ orbitals. For the description of
the dynamics of the valence
proton with respect to the core nucleus in $^8$B,
we can use the one particle density 
operator.  The radial dependence of the one body density operator,
in the single particle operator approximation, can be written as:
\begin{equation}
\rho_{j}(r) = \sum_{jm} a^{\dag}_{jm}(r)a_{jm}(r),
\end{equation}
where $a^{\dag}(r)[a(r)]$ is the creation (annihilation) operator for
nucleons in the radial coordinate $r$ and in single particle state
$\vert nljm\rangle$. The above radial density can be transformed
from particle operator representation to quasiparticle operator basis.
This leads to the HF-BCS vacuum expectation value as,
\begin{equation}
\rho(r) = \sum_{j}2\Omega_{j}v^{2}_j\phi^{2}_{j}(r),
\end{equation}
where $\Omega_{j} = j_{y} + 1/2$ and $\phi(r)$ are 
radial single particle SkHF wave functions.
Eq. (22) is the conventional expression for the one body
radial density function in the quasiparticle  representation.
 
In fig. 2, we show a comparison of the density distribution for
the valence proton and that of protons in $^8$B 
and $^7$Be nuclei. These densities are described above in
terms of the quasi particle operator representation. As can be seen from
this figure, the valence proton and $^8$B proton densities are similar
for distances beyond the rms radius of $^7$Be (2.50 fm).
In order to exhibit clearly the comparison of densities at 
10 fm, the density distributions are also shown on the linear scale
in the region 9-11 fm in inset of this figure.

In table IV, we also
show the theoretical results of spectroscopic factors
(which can be calculated by integrating the square of the norm of 
quasi particle HF+BCS radial wave functions or from the orbital occupancies) 
and astrophysical $S_{17}$ factors obtained by using Eq (20) using both
TH1 and TH2 force parameters.
Values of the spectroscopic factors for 
$p_{3/2}$ and $p_{1/2}$ orbitals do not differ significantly from
each other in the two cases.  However, the valence proton density at 10 fm
is lower in SkHF with TH2 force case by a factor of about 2 
as compared to that with TH1 force. This leads to a considerable
lower value for the $S_{17}$ factor in the former case.  
With TH1 density at 10 fm  
we get the astrophysical $S_{17}$ factor of 
$19.7$ eV b which is about the same as that obtained above. 
However, in the potential  model calculations Ref. \cite{bro96}
the difference between $S^A_{17}$ and $S^B_{17}$ is quite large.   

\section{Summary and conclusions}

In summary, in this paper we studied the structure of $^8$Li, $^8$B,
$^7$B, $^7$Be, and $^9$C nuclei within the Skyrme Hartree-Fock 
framework with $SLy4$ interaction parameters whose spin-orbit part is 
renormalized so as to reproduce the last proton binding energy in $^8$B.
The adjusted spin orbit term is weaker than that of the original force.
We calculated binding energies, various densities distribution and rms radii
for these nuclei. Using the same set of the force parameters, we obtain 
good agreements with experimental values of binding energies and
rms matter radii for all these nuclei. We have calculated the overlap
function $<^7$Be$\mid ^8$B$>$ from the SkHF wave functions which has been
employed to obtain the radius of the valence proton
in $^8$B nucleus. The value of this quantity is found to be  
4.93 fm which is almost
two times larger than the  matter radius of $^7$Be core. This 
provides support to the possibility of $^8$B having a one proton halo 
structure.

The same overlap function is used to extract the
asymptotic normalization coefficients for $^8$B $\rightarrow ^7$Be $+p$.
Using the overlap function calculated with the modified force we 
obtain an astrophysical S-factor of 22.0 eV b while the original
parameterization leads to  a value of 35.0 eV b. Thus the $S_{17}$
calculated with the TH1 force lies within the 
adopted limits ($19.1^{+4.0}_{-1.0}$ eV b) of the near zero
energy astrophysical S-factor. The values of $S_{17}$ obtained by using the 
corresponding asymptotic normalization coefficients follow the similar
trend. Although some dependence on the scattering model used in 
these calculations is apparent.  
We also calculated $S_{17}$ from the valence proton density
in the asymptotic region and the spectroscopic factors of the p-orbitals in
$^8$B. The value of the $S_{17}$ factor obtained in this way is 19.7 eV b,
which is quite close to the value obtained from the overlap function.
The average of these two values is 21.1 eV.b.   

The results of our calculations have a strong dependence on the 
parameter of the spin-orbit term of the Skyrme interaction. This 
suggests that it may be possible to have some important clue about 
the effective interaction in drip line nuclei from the comparison
of calculations with some experimental observables. 
\section{ Acknowledgment}

This work is supported by the Department of Atomic Energy,
(BRNS), (BARC), Mumbai, under contract no. 2001/37/14/BRNS/699.

\newpage
\begin{table}
\caption{The comparison of theoretical binding energies for various
nuclei calculated in self-consistent SkHF method with  experimental
data. The TH1 and TH2 represent theoretical results obtained 
with modified and original values of the $(b_4)$ parameter of
the SLy4 Skyrme force, respectively.} 
\begin{ruledtabular}
\begin{tabular}{cccc}
\multicolumn{1}{c}{Nucleus} & \multicolumn{3}{c}{BE(MeV)} \\
\hline
 &\multicolumn{1}{c}{ Expt.} & \multicolumn{2}{c}{Theor.}  \\
 & & TH1 & TH2  \\
\hline
$^{7}Be$ & $37.601$ & $37.561$&39.780   \\
$^{7}B$ & $24.720$ & $24.262$&25.931  \\
$^{8}Li$ & $41.278$ & $41.034$& 44.098  \\
$^{8}B$ & $37.739$ & $37.739$&40.677  \\
$^{9}C$ & $39.716$ & $39.035$&37.416  \\
\end{tabular}
\end{ruledtabular}
\end{table}
\newpage
\begin{table*}
\caption{ Rms mass ($r_m$), proton ($r_p$) and neutron ($r_n$) radii for
various nuclei.  
The TH1 and TH2 represents the theoretical results obtained with
modified and original values of the $(b_4)$ parameter of the SLy4 Skyrme
force, respectively. The experimental values of rms radii in $^7$Be [6],
$^8$Li [16], $^8$B [7,16] and $^9$C [38] are also shown here.}
\begin{ruledtabular}
\begin{tabular}{cccccccccc}
Nucl. & \multicolumn{9}{c}{rms radii(fm)}\\ \hline
    & \multicolumn{3}{c}{Expt.} & \multicolumn{6}{c}{Theor.}\\
&    & &  & \multicolumn{3}{c}{TH1}&
     \multicolumn{3}{c}{TH2}\\
\hline
 & $r_{m}^{exp}$ & $r_{p}$ & $r_{n}$ & $r_{m}$
 & $r_{p}$ & $r_{n}$ & $r_{m}$ & $r_{p}$
 & $r_{n}$\\
\hline
 $^{7}Be$ &$2.33\pm0.02$ & - & - &2.49 &2.63 &2.29 &2.32 &2.46 &2.12 \\
 $^{7}B$  &-  &- &- &2.86 &3.18 &1.84 &2.73 &3.01 &1.87 \\
 $^{8}Li$ &$2.37\pm0.02$ &$2.26\pm0.02$ &$2.44\pm0.02$ &2.54 &2.29 &2.67 &3.01 &2.98 &3.02 \\
 $^{8}B$  &$2.55\pm0.08$ &$2.76\pm0.08$ &$2.16\pm0.08$ &2.57 &2.73 &2.27 &2.84 &2.96 &2.73 \\
   &$2.43\pm0.03$ &$2.49\pm0.03$ &$2.33\pm0.03$ & & & & & & \\
 $^{9}C$  &$2.42\pm0.03$ & & &2.59 &2.77 &2.20 &2.13 &2.32 &1.67 \\
\end{tabular}
\end{ruledtabular}
\end{table*}
\newpage
\begin{table}
\caption{ SkHF results for the root-mean-square radius $<r^2_{^8B}>^{1/2}$
for the  $^7$Be + p = $^8$B bound state and rms matter radius $<r_m^2>^{1/2}$
in unit of fm. 
}
\begin{ruledtabular}
\begin{tabular}{cccc}
MODEL(ref.) & $<r^2_{^8B}>^{1/2}$ & $<r_{m}^2>^{1/2}_A$ & Ref. \\ \hline
            &                     &                   & \\
SkHF & $4.93$ & $2.74$ & present work \\
RPA+Mean-field & $4.73$ & $2.70$ & [10,11] \\
Microscopic & $4.43$ & $2.68$ & [12] \\
many-body &  &  &  \\
calculation &  &  &  \\
Two-body &$3.75$  &$2.52$  & [37] \\
Model &  &  &  \\
Cluster Model & $-$ & $2.50\pm 0.04$ & [15] \\
ANC method & $4.20\pm0.22$ & $2.60\pm0.04$ & $[13]$ \\
Exp. & $4.64\pm0.23$ & $2.83\pm0.06$ & [8,9] \\
Exp. & $3.97\pm0.12$ & $2.55\pm0.18$ & [7] \\
Exp. & $4.22$-$4.50$ & $2.58-2.60$ & [14] \\ 
Exp. & $-$ & $2.72$ & [16] \\ 
\end{tabular}
\end{ruledtabular}
\end{table}
\newpage
\begin{table}
\caption{SkHF results for    
asymptotic normalization coefficients
($\bar{c}_{l,j}$), spectroscopic factors ($S_{lj_y}$),
densities of valence particle ($\rho$) at 10 fm radius, 
and astrophysical $S-$ factors $S^A_{17}$ and $S^B_{17}$.
$S_{17}^{A1}$ corresponds to results obtained by using the SkHF
overlap function directly to the capture code
while $S_{17}^{A2}$ and $S_{17}^{A3}$ to those
obtained by using Eq. (19) with 
$C_N$ = 36.5 and 37.8, respectively.}
\begin{ruledtabular}
\begin{tabular}{cccc}
& \multicolumn{2}{c}{SkHF} \\
 & TH1 &TH2 \\ \hline
$\bar{c}_{1,3/2}$ & $0.64$ &$0.88$ \\
$\bar{c}_{1,1/2}$ & $0.34$&$0.29$  \\
$S^{A1}_{17}(eV b)$ & $22.0$ &$35.3$\\
$S^{A2}_{17}(eV b)$ & $19.5$ &$31.3$\\
$S^{A3}_{17}(eV b)$ & $20.2$ &$32.4$\\
\\ \hline
$S_{0p_{3/2}}$ & $1.02$ & $0.95$   \\
$S_{0p_{1/2}}$ & $0.42$ & $0.45$ \\
$\rho(10fm)\times 10^{-6}$ & $4.58$ \\
$S^B_{17}(eV b)$ &$19.7$ &$11.9$  \\
\end{tabular}
\end{ruledtabular}
\end{table}

\begin{figure}
\begin{center}
\includegraphics{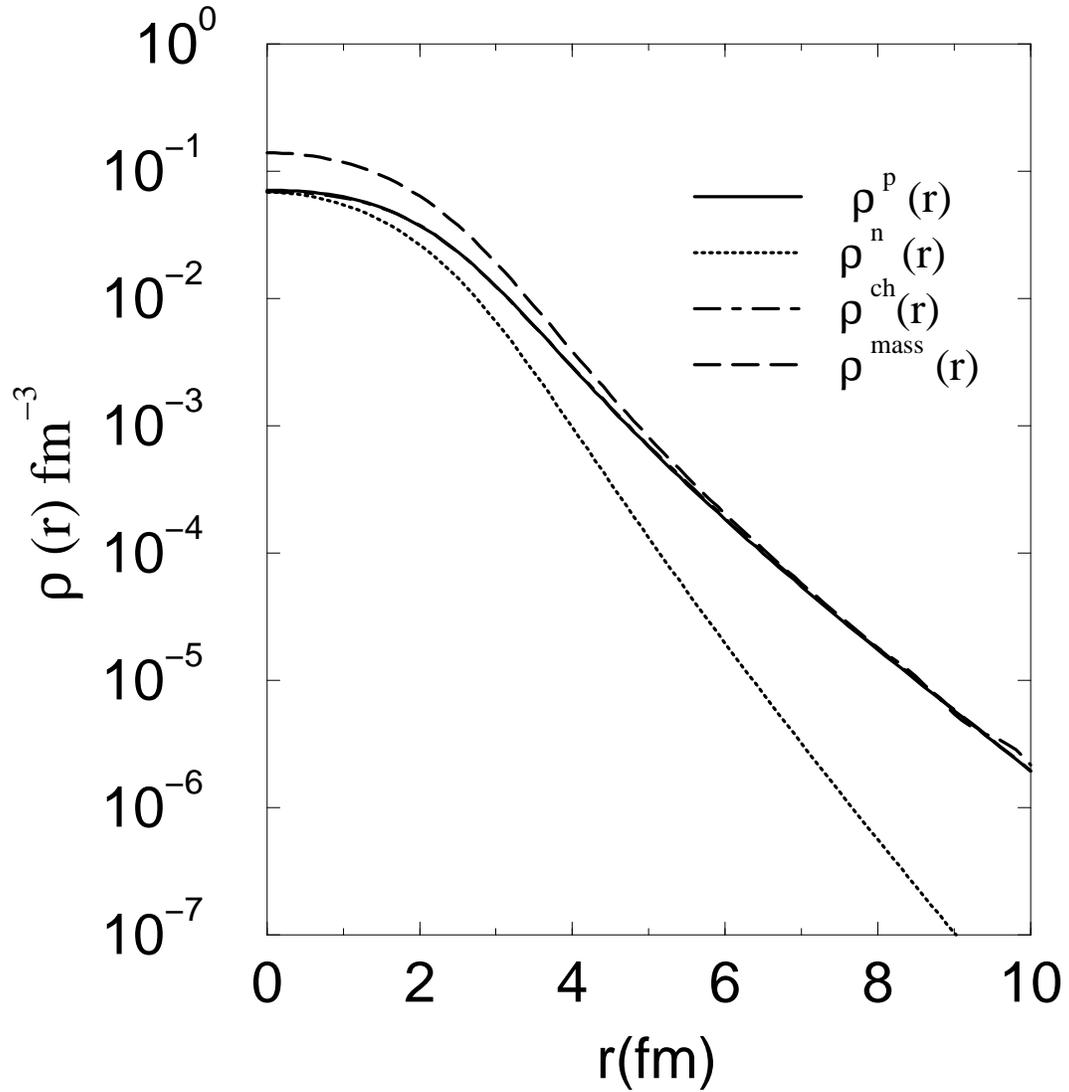}
\end{center}
\vskip .1in
\caption{Density distribution $\rho(r)$ for protons, neutrons,
nuclear charge and mass  in $^{8}B$ nucleus calculated with SkHF method.} 
\end{figure}
\newpage
\begin{figure}
\begin{center}
\includegraphics{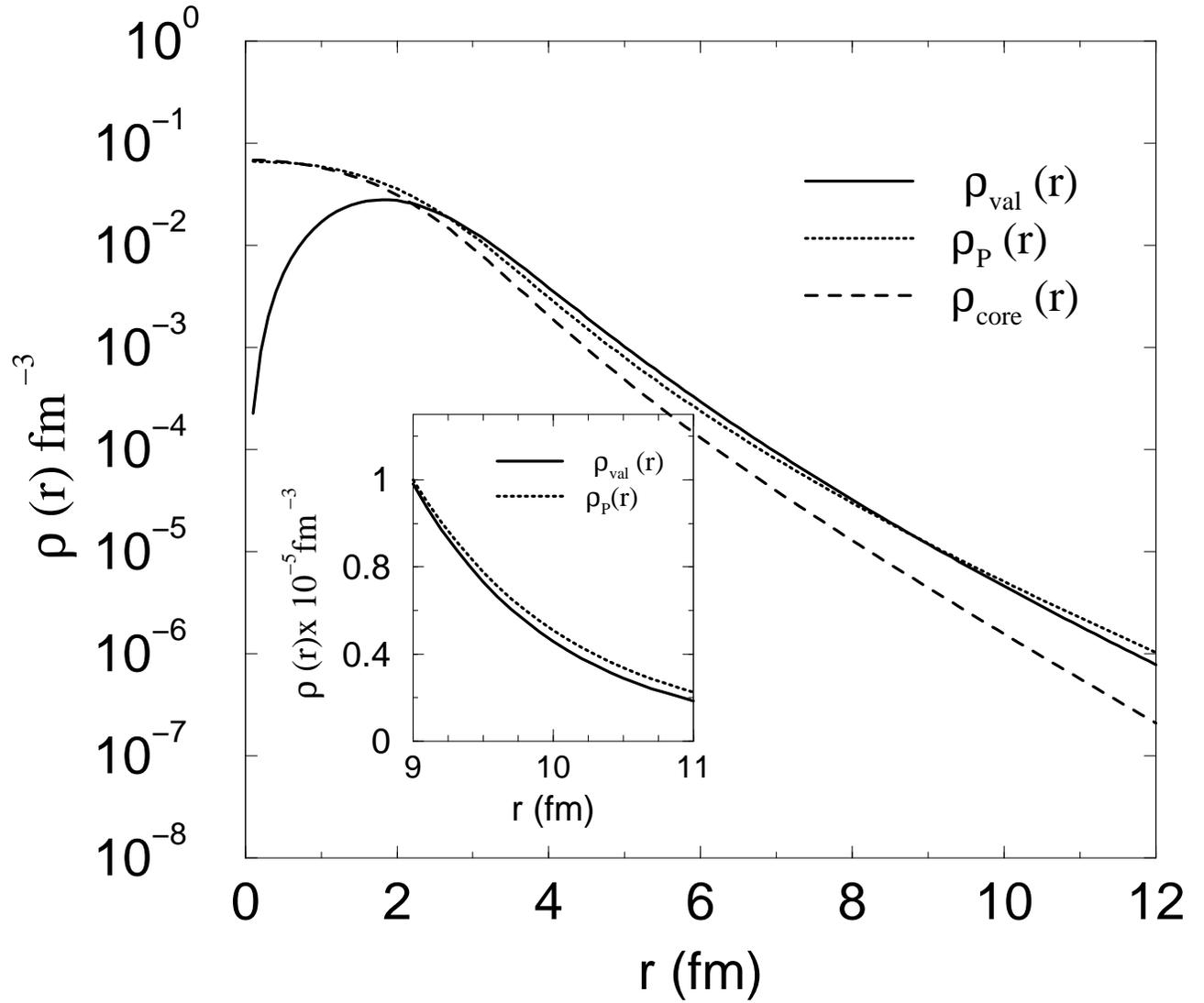}
\end{center}
\vskip .1in
\caption{The valence proton density $\rho_{val}(r)$ and the
proton density $\rho_{p}(r)$ in $^{8}B$ as a function of the radial
coordiante. Also shown is the r-space distribution of   
the density of core nucleus, $\rho_{core}(r)$. 
}
\end{figure}

\end{document}